\documentclass[review]{elsarticle}
\usepackage{geometry}
\newgeometry{left=2cm,right=2cm,top=2cm,bottom=3cm}
\usepackage{lineno,hyperref}
\modulolinenumbers[5]
\usepackage{amsmath}
\usepackage{amssymb}
\usepackage{cases}
\usepackage{float}

\journal{Journal of \LaTeX\ Templates}

\begin{document}

\begin{frontmatter}

\title{Study on alpha decay chains of $Z=122$ superheavy nuclei with deformation effects and Langer modification}

%% Group authors per affiliation:

%% or include affiliations in footnotes:

\author[mymainaddress]{G. Naveya}

\author[mysecondaryaddress]{S. I. A. Philominraj}

\author[mymainaddress]{A. Stephen\corref{mycorrespondingauthor}}
\cortext[mycorrespondingauthor]{Corresponding author}
\ead{stephen_arum@hotmail.com }

\address[mymainaddress]{Department of Nuclear Physics, University of Madras, Chennai-600025, India}
\address[mysecondaryaddress]{Department of Physics, Madras Christian College, Chennai-600059, India}

\begin{abstract}
In this work study on alpha decay chains emerging from isotopes of $Z=122$ superheavy nuclei is carried out with emphasize on nuclear deformations and Langer modification. The interest in this particular superheavy nuclei is due to the recent experimental efforts to synthesize the isotope $^{299}{120}$ in a fusion reaction at the velocity filter SHIP (GSI Darmstadt), which makes synthesis of $Z=122$ nuclei to occur in the near future, and in turn will give the experimentalist the chance observe the decays associated with the isotopes of this nuclei. We perform our calculations by choosing the Woods Saxon potential for nuclear interaction, along with Coulomb potential and centrifugal potential within the framework of the WKB method.  
When the centrifugal term is taken in the total potential and WKB integral is done over 1D radial coordinate, it requires the use of Langer modification wherein $(l+ \frac{1}{2})^{2}$ replaces $l(l+1)$ for consistency of WKB wave function. Hence we have used this Langer modified centrifugal potential. The orientations of deformed nuclei are important, as it affects the touching distance and also influence the nuclear and Coulomb potential, and thus can alter the values of penetration integral and half-life.  The results obtained by our calculations are in agreement with the values obtained from various phenomenological models. This study brings out the unique characteristics of alpha chains associated with each isotope, which will be helpful for the experimentalist to decide the isotope they would like to synthesize and also for their post-synthesis study.
\end{abstract}

\begin{keyword}
Alpha decay \sep Woods-Saxon \sep Deformation  \sep Half-lives \sep Langer Modification \sep Spontaneous fission
\end{keyword}

\end{frontmatter}

\section{Introduction}
In 1911 Ernest Rutherford performed his gold foil experiment which leads to the discovery of the atomic nucleus, which was thought to be a small dense region consisting of protons at the center of an atom. After the discovery of the neutron in 1932, shell models for nucleus consisting of protons and neutrons were developed by Dmitri Iwanenko and Werner Heisenberg \cite{iwanenko1932} \cite{heisenberg1932}. Over the years both in theoretical and experimental front, the atomic nucleus has been studied extensively. Elements up to Uranium 92 can be found in nature, the majority of these elements consist the light and medium-heavy nuclei, with those above Z $>$ 83 posses heavy nuclei. Elements beyond U-92 are called as transuranium elements and experimentalist have synthesized them in the laboratory and they offer us with heavy and superheavy nuclei. A superheavy element, in general, is referred to elements with an atomic number greater than 104. The laboratories involved in the discovery of the transuranium element are : Lawrence Berkeley National Laboratory in the United States (Z= 93-101, 106, and Z= 103-105), the Joint Institute for Nuclear Research in Russia (Z= 102 and Z=114-118, and Z=103-105), the GSI Helmholtz Centre for Heavy Ion Research in Germany (Z=107-112), and RIKEN in Japan (Z=113) \cite{elementdata}. The heaviest element known so far is Z = 118. The nuclei of the elements present in our periodic table and their isotopes exhibit diverse neutron-proton ratio, binding energy, size, and stability.

Radioactive decay is an important phenomenon associated with the nucleus, in which a nucleus undergoes transition via decay modes which can be alpha decay, beta decay, gamma decay, neutron emission, proton emission, spontaneous fission, and cluster decay. Light and medium-heavy nuclei decay almost entirely via beta decay, electron capture, and proton emission. Heavy and superheavy nuclei decay can decay via beta decay, alpha decay, and spontaneous fission, but beta decay for the superheavy nuclei is slow as it proceeds via a weak interaction so is less favored compared to spontaneous fission and alpha decay. Cluster radioactivity is a rare process where nuclei decay and emit a fragment which is heavier than the alpha particle but lighter than the fission fragment. The theoretical prediction of such phenomena was put forward by Sandulescu, Poenaru, and Greiner in 1980 \cite{sand1980}, within few years in 1984, Rose and Jones experimentally observed this radioactivity where $^{223}Ra$ emitted  $^{14}C$ \cite{rose1984}. In last three decades many other heavier clusters have been observed which include $^{20}O$, $^{23}F$, $^{24-26}Ne$, $^{28-30}Mg$, and $^{32}Si$ \cite{bonetti2007} \cite{dongdong2010}, where most of the parent turns out to be a heavy nuclei. The cluster decay from superheavy nuclei is yet to be observed experimentally.

Theoretical studies have been performed on both the cluster decay and alpha decay of heavy and SH nuclei within various theoretical models, some of which are Generalized liquid drop model (GLDM) \cite{royer2001}, preformed cluster model (PCM) \cite{malik1989}, Coulomb Proximity Potential model( CPPM) \cite{kpsantosh2012}, Yukawa plus exponential (Y+E) potential model \cite{shanmugam1995}, and Analytical Super-Asymmetric Fission model (ASAF) \cite{poenaru2018b}. Various half-life phenomenological formula's for alpha decay and cluster decay exist in literature such as Viola- Seaborg formula \cite{viola1966}, universal decay law \cite{qi2009}, Royer's formula \cite{royer2010}, universal curve \cite{poenaru2011}, AKRA \cite{akrawy2017}. In recent years the concept of cluster radioactivity has beem reframed to accomodate the emission of particles with the charge number $Z_{c}> 28$ from the parent nuclei with $Z_{p}= Z _{c} + Z_{d} > 110$\cite{wei2017}. Poenaru and Gherghescu theoretically investigated the $^{92,94}{Sr}$ cluster radioactivity of $^{300,302}{120}$  and predicted a branching ratio relative to alpha decay being $−0.10$ and $0.49$ respectively, which suggests that such cluster decay modes have could be observed\cite{poenaru2018}. Latest experiment at the velocity filter SHIP (GSI Darmstadt) aimed to produce the $^{299}{120}$ isotope in a fusion reaction $^{248}{Cm}(^{54}{Cr},3n)$ $^{299}{120}$ \cite{hofmann2016}. Hence in near future with the synthesis of $^{300,302}{120}$ isotopes large clusters like $^{92,94}{Sr}$ can be expected to be observed in the decay.

The synthesis of $Z=122$ nuclei isotopes will take place in the near future. We do our theoretical study within the framework of the WKB method to investigate the emission of alpha chain from the even isotopes of $Z=122$ by taking nuclear deformation and orientation affects into consideration. Each isotope has a unique alpha chain signature associated with it, we aim to theoretically predict the features of alpha chain from superheavy nuclei having $198 \leq A \leq 316$ with $Z=122$. Superheavy nuclei can decay either via alpha decay or via spontaneous fission. In spontaneous fission, they become split into too big fragments. The alpha emission in the form of chains from a particular superheavy nucleus continues until spontaneous fission is encountered. Hence the spontaneous fission half-lives are studied along with alpha decay half-lives for the decay chains of each isotope of $Z=122$, restricting our calculation of spontaneous fission using the phenomenological formula.
 
\section{The Model}

During alpha decay or cluster decay, a parent nucleus gets split into a daughter nuclei and a fragment. The interaction between daughter and fragment is majorly influenced by the nuclear potential and Coulomb potential with some contribution also from the centrifugal potential. In general, the daughter and fragment nuclei are deformed and can have a certain orientation. Consider $\theta_{1}$ and $\theta_{2}$ to be the angle between the radius vector and axis of symmetry, which determines the orientation of daughter and fragment respectively.

\begin{figure}[H]
\center
\includegraphics[height=5cm, width=9cm]{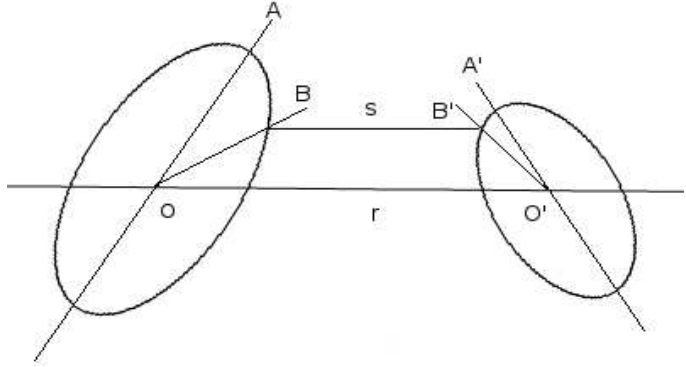}
\caption{Daughter and fragment are shown, r is the between their center. s is the shortest distance between the nuclei. OA is the axis of symmetry of daughter and O'A' is the axis of symmetry for the fragment. $\angle  AOB= \theta_{1}$ and  $ \angle A'O'B'=\theta_{2}$.}
\end{figure}

The orientations of deformed nuclei are important, as it affects the touching distance and also influences the nuclear and Coulomb potential, and thus on the values of penetration integral and half-life. Hence it is essential to consider a total potential which is dependent on the orientation angles of deformed nuclei. The total potential is taken as,

\begin{equation}
V(r,\theta_{1},\theta_{2})=V_{n}(r,\theta_{1},\theta_{2} ) + V_{c} (r,\theta_{1},\theta_{2} ) + V_{l}(r)
 \end{equation}

To describe the nuclear interaction Woods-Saxon potential is taken. The form of potential is the one suggested by Sandulescu et al. \cite{sand1980} \cite{buck1996}. It is given as

\begin{equation}
V_{n}(r,\theta_{1},\theta_{2})= - \frac{V_{0}}{1+ \exp{\left[\frac{r-R(\theta_{1},\theta_{2})}{a}\right]}}
\end{equation} 

$V_{0}$ is the depth of nuclear potential which is taken to be

\begin{equation}
V_{0}=18(A_{1}^{2/3}+A_{2}^{2/3}-(A_{1}+A_{2})^{2/3})
\end{equation}

$A_{1}$ is the atomic number of the daughter and $A_{2}$ is the atomic number of cluster. The diffuseness parameter $a$ can be calculated using 

\begin{equation}
a=0.5+0.33(I_{2}+I_{1})
\end{equation}

where $I_{i} =\frac{N_{i} - Z_{i}}{A_{i}}$ with $N_{i}$ being the number of neutrons present in the nuclei \cite{rahimi2018}. The radius of daughter and fragment can be found using

\begin{equation}
R_{0i}=(1+0.39 I_{i})A_{i}^{1/3}
\end{equation}

Taking the deformation and orientattion into consideraion the effective nuclear radius of daughter or fragment can be evaluated \cite{kpsantosh2018} \cite{malhotra1985}. A relaible formula present in literature is,

\begin{equation}
R_{i}(\theta_{i})= R_{0i} \left[ 1 + \sum_{\lambda=2,3,4,6} \beta_{\lambda i} \hspace{0.1cm} Y_{\lambda}^{0} (\theta_{i}) \right]
\end{equation}

where $\beta$'s are deformation parameters and  $Y_{l}^{m}$ are the spherical harmonics. The orientations of the daughter and parent nuclei along with their deformations determine the touching distance. For ease, the touching distance is assumed to be the same as the effective nuclear radius R of the Woods-Saxon potential. The effective nuclear radius is calculated using the expression in ref\cite{rahimi2018}, which is

\begin{equation}
R(\theta_{1},\theta_{2})=1.17 + R_{1}(\theta_{1}) + R_{2}(\theta_{2})
\end{equation}

The nuclear potential is attractive in nature and it mimics the interaction between daughter and fragment when they are within the parent (for $r < R$). The Coulomb potential is repulsive in nature. Some works on decay studies consider a very simple form of the Coulomb potential, where the repulsive Coulomb potential acts between fragment and daughter only in the region $r \geq R$  \cite{saidi2015}, it is given by 

\begin{numcases}{V_{c}(r)=}
0,& r $\leq$ R \\
\frac{Z_{1} Z_{2} e^{2}}{r} ,& r $\geq$  R
\end{numcases}

This is similar to the one considered by Gamow to study the alpha decay. Whereas in some works, the authors have considered Coulomb interaction between the daughter and the fragment to be present when they are inside the parent ($r \leq R$)  and also when they come out of the parent ($r \geq R$). With an assumption that the fragment is a point charge which is present within a  uniformly charged sphere \cite{sand1992}, the Coulomb potential comes out to be

\begin{numcases}{V_{c}(r)=}
 \frac{Z_{1} Z_{2} e^{2}}{2R}\left(3 -\frac{r^{2}}{R^{2}} \right),& r $\leq$ R \\
\frac{Z_{1} Z_{2} e^{2}}{r} ,& r $\geq$  R
\end{numcases}

Some authors have studied the deformations of daughter and fragment together with their individual orientations. The Coulomb potential is taken in such works acts between fragment and daughter only after it comes out of parent($r \geq R$)   \cite{rahimi2018} \cite{kpsantosh2018}, it is given by

\begin{numcases}{V_{c}(r)=}
0,& r $\leq$ R \\
\frac{Z_{1} Z_{2} e^{2}}{r} + \frac{3 Z_{1} Z_{2} e^{2}}{r} \sum_{i=1,2}\left[ \sum_{\lambda=2,3,4,6} \left \{ \hspace{0.1cm} \frac{R_{0i}^{\lambda}}{r^{\lambda}} \hspace{0.1cm} \times\hspace{0.1cm}  \beta_{\lambda} \hspace{0.1cm} Y_{\lambda}^{0} (\theta_{i}) \hspace{0.1cm} \times \left(1+ \frac{4}{7}\beta_{\lambda}\hspace{0.1cm} Y_{\lambda}^{0} (\theta_{i}) \delta_{\lambda,2} \right)  \right \}  \right] , & r $\geq$  R
\end{numcases}

In this work, the Coulomb interaction between daughter and fragment is assumed to be present even when they are inside the parent ($r \leq R$) and also when they come out of parent ($r \geq R$), additionally the deformation and orientation are taken. This is to mimic the realistic scenario which may be happening during the decay. The potential used in this work is

\begin{numcases}{V_{c}(r)}
\approx  \frac{Z_{1} Z_{2} e^{2}}{2R}\left(3 -\frac{r^{2}}{R^{2}} \right),& r $\leq$ R \\
=\frac{Z_{1} Z_{2} e^{2}}{r} + \frac{3 Z_{1} Z_{2} e^{2}}{r} \sum_{i=1,2}\left[ \sum_{\lambda=2,3,4,6} \left \{ \hspace{0.1cm} \frac{R_{0i}^{\lambda}}{r^{\lambda}} \hspace{0.1cm} \times\hspace{0.1cm}  \beta_{\lambda} \hspace{0.1cm} Y_{\lambda}^{0} (\theta_{i}) \hspace{0.1cm} \times \left(1+ \frac{4}{7}\beta_{\lambda} \hspace{0.1cm} Y_{\lambda}^{0} (\theta_{i}) \delta_{\lambda,2}  \right)  \right \}  \right] , & r $\geq$  R
\end{numcases}

It is to be noted that when $r \leq R$, the Coulomb potential is taken is approximately the one as experienced by a test charge present at r within a uniformly filled sphere of radius R. This can be a good approximation when our fragment is an alpha particle or small cluster. It is difficult to consider the deformation and orientation effects of both daughter and fragment when they are within the parent nuclei ($r \leq R$), hence have approximated it. The Coulomb potential between the daughter and the fragment when they are at $r \geq R$ is expressed in terms of their deformation and also on their orientation.\\

The centrifugal potential term depends on the angular momentum quantum number $l$. Most works have taken the centrifugal potential to be

\begin{equation}
V_{l}(r) = \frac{l(l+1) \hbar^{2}}{2 \mu r^{2}}
\end{equation}

where $\mu$ is the reduced mass of daughter and fragment system.\\

Using this form of centrifugal potential has an issue associated with it when used in  1D WKB approximation. It is well known that the radial part of Schrodinger equation in 3D can be written in the same form as  Schrodinger equation in 1D cartesian coordinates, but applying the WKB method for radial coordinate to derive the energy eigenvalues is not always correct. For example, it has been found that using the first-order WKB integral the eigenvalues of a hydrogen atom and harmonic oscillator obtained, to be consistent the quantity $l(l+1)$ has to be replaced by $(l+\frac{1}{2})^{2}$ in the quantization rule \cite{langer1937} \cite{gu2008}. Langer (1937) showed that such a substitution is necessary for correct behavior of WKB wave function near the origin and also for the validity of the connection formula used. Hence in this work, the Langer modified centrifugal potential term is considered, given by

\begin{equation}
V_{l}(r) = \frac{(l+\frac{1}{2})^{2} \hbar^{2}}{2 \mu r^{2}}
\end{equation}

Langer modified centrifugal potential has been previously employed in some of the decay related works\cite{soylu2015} \cite{ismail2017}. The classical turning point is obtained by solving for $V(r)- Q=0$, using the Q value of decay process. There exists 3 turning points $r_{1}(\theta_{1},\theta_{2})$, $r_{2}(\theta_{1},\theta_{2})$, and $r_{3}(\theta_{1},\theta_{2})$. A barrier exists between inner turning point $r_{2}(\theta_{1},\theta_{2})$ and outermost turning point $r_{3}(\theta_{1},\theta_{2})$. Crossing this barrier the fragment quantum mechanically tunnels through.\\

The barrier penetrability can be obtained using WKB method,
\begin{equation}
P(\theta_{1},\theta_{2}) = \exp[-2K(\theta_{1},\theta_{2})]
\end{equation}
with 
\begin{equation}
K(\theta_{1},\theta_{2})= \frac{1}{\hbar} \int_{r_{2}(r,\theta_{1},\theta_{2})}^{r_{3}(\theta_{1},\theta_{2})} \sqrt{2 \mu (V(r)-Q)} \hspace{0.1cm} dr
\end{equation}

Taking the different combination of orientations of daughter and fragment $\theta_{1}$ and $\theta_{2}$, can find the P for each combination of angles. By averaging the mean value of barrier penetrability can be calculated.

\begin{equation}
P_{average} = \frac{1}{n} \sum_{ \{(\theta_{1},\theta_{2}) \}} P(\theta_{1},\theta_{2}) 
\end{equation}
where n is the total number of combinations of $(\theta_{1},\theta_{2})$ considered. To perfrom our calculations 400 combinations of ($\theta_{1},\theta_{2}$) is used, such that $ 0 \leq \theta_{1} \leq \pi $ and  $ 0 \leq \theta_{2} \leq \pi $. The decay constant is given by \cite{sand1992}, 
\begin{equation}
\lambda= P S \nu 
\end{equation}

where $\nu$ is the assault frequency and $S$ is the spectroscopic facor which describes the preformation probability. The assualt frequency can be obtained using \cite{kpsantosh2018} \cite{rahimi2018},

\begin{equation}
\nu = \frac{2 E_{v}}{h}
\end{equation}

where $E_{v}$ is the vibrational energy.  $E_{v}$ is taken to be Q value of the decay process to carry out the calculation. The spectroscopic factor for alpha particle emission is, $S_{\alpha} = 6.3 \times 10^{-3}$ for even mass parent nuclei and $S_{\alpha} = 2.3 \times 10^{-3}$ for odd mass parent nuclei. For the cluster decay  $S_{c} = S_{\alpha}^{(A_{2}-1)/3}$ \cite{sand1980}. Half-life can be calculated using
  
\begin{equation}
T_{1/2}=\frac{\ln2}{\lambda}
\end{equation}

Most works have restricted themselves to the use of only $\beta_{2}$ and $\beta_{4}$ \cite{soylu2015} \cite{rahimi2018}, and additionally some authors have used also the $\beta_{3}$ \cite{kpsantosh2018}. But $\beta_{6}$ is non zero for several nuclei and hence can have some effect on the touching distance, barrier penetrability and hence on half-lives. The deformation parameter values  $\beta_{2}$, $\beta_{3}$, $\beta_{4}$ and $\beta_{6}$ are used for our calculation. From the table of Moller et. al, the deformation parameters are taken and the Q value is evaluated using the mass excess data \cite{moller2016}. In this study calculations are done for $l=0$, the Langer modified centrifugal term is non-zero for $l=0$ which differs from traditional centrifugal term, and it contributes to the value of penetration integral.\\

\section{Phenomenological formulas: For alpha decay and spontaneous fission}

Various phenomenological formulas exist in the literature for alpha decay and spontaneous fission. With the phenomenological formula's described in this section, the alpha decay half-lives are evaluated. This is to check if our model can give half-life values which are close to phenomenological formulas.

\subsection{\bf Universal Decay Law}

The universal decay law (UDL) for alpha decay and cluster decay gives the half life using the relation,

\begin{equation}
log_{10}T_{1/2}= a X^{'} + B \rho' + c
\end{equation}

where  $$X^{'}= Z_{1} Z_{2} \sqrt{\frac{A}{Q}} \hspace{0.1cm}, \hspace{0.5cm}  \rho^{'}= \sqrt{A Z_{1} Z_{2} (A_{1}^{1/3}+A_{2}^{1/3})} \hspace{0.1cm},\hspace{0.5cm}   A=\frac{A_{1}A_{2}}{A_{1}+A_{2}}$$.\\ 

C. Qi et al., have obtained the values of coefficients a,b, and c by fitting this relation to experimental data. By fitting to experimental data of alpha decay and cluster decay half-lives together it has been reported that, $a= 0.4314, b=-0.4087, c=-25.7725$  \cite{qi2009}.\\

\subsection{\bf Viola-Seaborg formula}
Viola-Seaborg relation is one of the widely used relations for calculating the alpha decay half-lives \cite{viola1966}. It is given by,

\begin{equation}
log_{10}T_{1/2} = (aZ + b)Q ^{−1/2} + cZ + d + h_{log}
\end{equation}

The values of coeficients a, b, c, d and $h_{log}$ we take from the work of T. Dong et. al. \cite{dong2005}, We use $a=1.64062 $   $b=-8.54399  $ , $ c=-0.19430 $, $d=-33.9054$ and

\begin{numcases}{h_{log}=}
  0, & Z even and A even \\
  0.8937,& Z even and A odd   \\
   0.5720,& Z odd and A even  \\
   0.9380,& Z odd and A odd  
    
\end{numcases}

\subsection{\bf AKRA}

AKRA model was constructed by modification of Royer's formula \cite{akrawy2017}.  The Royer's formula is,

\begin{equation}
T_{1/2} = a+ b A^{1/6} \sqrt{Z} + c \frac{Z}{\sqrt{Q}}
\end{equation}

With adding new paramerts expression of following form was used,

\begin{equation}
T_{1/2} = a+ b A^{1/6} \sqrt{Z} + c \frac{Z}{\sqrt{Q}}+ d I + e I^{2}
\end{equation}

where $I= N-Z/A $. Poenaru et. al, have obtained the coefficients by fitting this relation to the experimental data. The values are $a = −27.989$, $b = −0.940$, $c = 1.532$, $d =−5.747$, and $e = 11.336$ \cite{poenaru2018}.

\subsection {\bf Spontaneous fission half life of Xu et. al}
A relaiable formula for half life of spontaneous fission is from the works of Xu et. al, which is of the form,

\begin{equation}
T_{1/2}= \exp\left[2\pi \left( c_{0} + c_{1}A +c_{2} Z_{2} + c_{3} Z_{4} + c_{4} (Z-N)^{2}- (0.13323 Z^{2}A^{-1/3} - 11.64) \right) \right]
\end{equation}

By fit with experimental data the values of coefficient reported by Xu et al., is $c_{0}= - 195.09227$, $c_{1} = 3.10156$, $c_{2} = -0.04386$, $c_{3} = 1.4030\times10^{-6}$ , and $c_{4} = 0.03199$ \cite{xu2008}.

\section{Result $\&$ Discussion}

\begin{table}[H]
\caption{Alpha decay chains from even istopes of $Z=122$}
\setlength\tabcolsep{3.8pt}
\begin{tabular}{cccccccccccccc}
\cline{1-13}
Parent & Daugter & \multicolumn{4}{c}{Daughter deformation} & $Q$ (MeV)  & $log_{10}T_{1/2}^{SF}$&\multicolumn{4}{c}{$log_{10}T_{1/2}^{\alpha}$}& Mode\\ \cline{3-6} \cline{9-12} 
&   & $\beta_{2}$ & $\beta_{3}$  &  $\beta_{4}$       & $\beta_{6}$ & &Xu& Model &  UDL & VS & AKRA\\ 
\cline{1-13}
$^{294}122$  & $^{290}120$  &   -0.125 & 0.000&  0.018& 0.008 & 14.515& 22.3886 & -7.40168 & -7.82952&-7.32498 &-6.36205 & $\alpha$\\
$^{290}120$  & $^{286}118$  & 0.075    &0.000 &0.014  &-0.009 &13.75 &15.0622&-6.79957 &-6.89695 &-6.43247 &-5.55713&$\alpha$\\
$^{286}118$  & $^{282}116$  &0.053    &0.000&0.025  &0.001  &13.05 & 9.16018& -5.9921&-6.03392 &-5.60781 &-4.81492&$\alpha$\\
$^{282}116$  & $^{278}114$  & 0.000   &0.000&0.000  &0.000  &12.62 & 4.59346& -5.41938 &-5.69865 &-5.27748 &-4.55238&$\alpha$\\
$^{278}114$  & $^{274}112$  & 0.221   &0.000&-0.093  &0.000  &11.76 & 1.27452&-4.65754 &-4.35052 &-4.00782 &-3.36772&$\alpha$\\
$^{274}112$  & $^{270}110$  & 0.232   &0.000  &-0.066  &-0.006  &11.41 &-0.882719 &-4.52782 &-4.13851 &-3.79837&-3.21668&$\alpha$\\
$^{270}110$  & $^{266}108$  & 0.232   &0.000&-0.052  &-0.023  &10.42 &-1.96295  & -2.73985&-1.5487 &-1.39026 &-0.858342& $\alpha$/ SF\\

\cline{1-13}

$^{296}122$  & $^{292}120$  &0.086   &0.000 &-0.034  &0.007 &14.95 &23.3441 &-8.36188  & -8.64387&-8.05342 &-7.11311 &$\alpha$\\
$^{292}120$  & $^{288}118$  &-0.086  &0.000 &-0.009  &-0.011 &13.77 &15.9907 & -6.59291 &-6.96907 &-6.46936 &-5.63356 &$\alpha$ \\
$^{288}118$  & $^{284}116$  &0.064  &0.000 &0.026  &-0.008 &12.86 &10.0614 &-5.68637& -5.66145&-5.23079 &-4.48691  &$\alpha$\\
$^{284}116$  & $^{280}114$  & 0.000   &0.000&0.000  &0.000  &12.22  &  5.46734 &-4.65049&-4.83856 &-4.4468 &-3.78041&$\alpha$\\
$^{280}114$  & $^{276}112$  &0.210    &0.000&-0.094  &0.001  &11.05  & 2.1209&-3.02491 & -2.61343 &-2.36173 &-1.79707&$\alpha$\\
$^{276}112$  & $^{272}110$  &0.221    &0.000&-0.080  &-0.007  &11.9  & -0.06400& -5.38574 &-5.33169 &-4.87748 &-4.31337& $\alpha$\\

\cline{1-13}

$^{298}122$  & $^{294}120$  &0.086  &0.000 &-0.046  &0.016 &15.16 &23.5852 &-8.76475& -9.04593&-8.39785 &-7.48859 &$\alpha$\\
$^{294}120$  & $^{290}118$  &0.086  &0.000 &-0.021  &-0.002 &13.49 &16.2047 & -6.37057   & -6.43871&-5.94536 &-5.16145  &$\alpha$\\
$^{290}118$  & $^{286}116$  &0.075  &0.000 &0.014  &-0.019 &12.68 &10.2483 &-5.33563   & -5.30171&-4.86582 &-4.16963 &$\alpha$\\
$^{286}116$  & $^{282}114$  &0.000    &0.000&0.000  &0.000  &11.68  &  5.62683&-3.50129 & -3.59376 &-3.25838 &-2.65764&$\alpha$\\
$^{282}114$  & $^{278}112$  &0.187    &0.000&-0.072  &-0.003  &9.96 &  2.25291&  -0.425281& 0.431024 &0.500072 &0.964215& $\alpha$\\

\cline{1-13}

$^{300}122$  & $^{296}120$  &0.075  &0.000 &-0.046  &0.017 &14.72 &23.1121 &-8.07904& -8.2944&-7.66776 &-6.81497 &$\alpha$\\
$^{296}120$  & $^{292}118$  &0.075  &0.000 &-0.034  &0.008 &13.59 &15.7046 & -6.57417 & -6.67381&-6.13436 &-5.38435  &$\alpha$\\
$^{292}118$  & $^{288}116$  &0.075  &0.000 &0.002  &-0.010 &12.39 &9.72094 & -4.77285   & -4.68428&-4.26116 &-3.6172 &$\alpha$\\
$^{288}116$  & $^{284}114$  &0.064    &0.000&0.014  &-0.009  &11.2 & 5.07221 &-2.63534 &-2.41387 &-2.13064 &-1.59316&$\alpha$\\
$^{284}114$  & $^{280}112$  &0.041    &0.000&-0.041  &-0.006  &9.52  & 1.67084 &1.27806  & 1.78839 &1.79227 &2.19069&$\alpha$/SF\\
\cline{1-13}
$^{302}122$  & $^{298}120$  &-0.063  &0.000 &0.002  &0.010 &14.77 &21.925&-8.03976 &-8.41722 &-7.75237 &-6.9351 &$\alpha$\\
$^{298}120$  & $^{294}118$  &0.064  &0.000 &-0.034  &0.008 &13.24 &14.490&-5.93156 & -5.98536&-5.46352 &-4.76695  &$\alpha$\\
$^{294}118$  & $^{290}116$  &0.064  &0.000 &-0.022  &-0.001 &12.37 &8.47973&-4.70893 &-4.67058 &-4.21868 &-3.61299 &$\alpha$\\
$^{290}116$  & $^{286}114$  & 0.064   &0.000&-0.010  &-0.001  &11.06 &3.80373& -2.287 &-2.07745 &-1.78797 &-1.29524&$\alpha$\\
$^{286}114$  & $^{282}112$  &0.086    &0.000&-0.009  &-0.011  &9.48 &0.374951& 1.47891 &1.88776 &1.91418 &2.27306& SF\\
\cline{1-13}

\cline{1-13}
\end{tabular}
\end{table}

In table 1, it can be seen that the decay chains for $^{296}{122}$, $^{298}{122}$, $^{300}{122}$ is discontinued abruptly at the 6th/5th step of chain process without the mode of decay being spontaneous fission reached. This is done because the $Q$ value calculated (using mass excess data from Moller table \cite{moller2016}) for next step of decay comes out to be negative, which indicates that the next step of alpha decay is endothermic and is not possible on own.  

\begin{table}[H]
\caption{Alpha decay chains from even istopes of $Z=122$}
\setlength\tabcolsep{3.8pt}
\begin{tabular}{ccccccccccccc}
\cline{1-13}
Parent & Daugter & \multicolumn{4}{c}{Daughter deformation} & $Q$ (MeV)  & $log_{10}T_{1/2}^{SF}$&\multicolumn{4}{c}{$log_{10}T_{1/2}^{\alpha}$}& Mode\\ \cline{3-6} \cline{9-12} 
&   & $\beta_{2}$ & $\beta_{3}$  &  $\beta_{4}$       & $\beta_{6}$ & &Xu& Model &  UDL & VS & AKRA\\ 
\cline{1-13}

$^{304}122$  & $^{300}120$  &-0.032  &0.000 &-0.011  &0.000 &14.59 &20.0243&-7.78426 & -8.11989&-7.44576 &-6.67271 &$\alpha$\\
$^{300}120$  & $^{296}118$  & -0.063 & 0.000&0.002  &0.010 &13.69 &12.5628&-6.70725 &-6.93834 &-6.32129 &-5.64079 &$\alpha$\\
$^{296}118$  & $^{292}116$  &-0.073  &0.000 &0.002  &-0.000 &12.28 &6.52489 & -4.38246 & -4.49558&-4.02623 &-3.46133 &$\alpha$\\
$^{292}116$  & $^{288}114$  &-0.021    &0.000&0.012  &-0.000  &10.82 &1.82165& -1.57293& -1.4612 &-1.18512 &-0.742007&$\alpha$\\
$^{288}114$  & $^{284}112$  & 0.086   &0.000&-0.021  &-0.002  &9.17 & -1.63451& 2.43706 &2.90107 &2.8859 &3.18793& SF\\
\cline{1-13}

$^{306}122$  & $^{302}120$  &0.000  &0.000 &0.000  &0.000 &14.61 &17.41&-7.9523 &-8.18842 &-7.4801 &-6.74201 &$\alpha$\\
$^{302}120$  & $^{298}118$  & -0.032 &0.000 &-0.011  &0.000 &13.56&9.9216&-6.52901 &-6.70858 &-6.07788 &-5.43881  &$\alpha$\\
$^{298}118$  & $^{294}116$  &-0.042  &0.000 &0.001  &0.000 &12.49&3.85665 &-4.92292& -5.00617&-4.47204 &-3.93228 &$\alpha$\\
$^{294}116$  & $^{290}114$  &-0.011    &0.000& 0.000 &0.000  &10.9 &-0.873791&-1.826  &-1.71106 &-1.38828 &-0.975868&$\alpha$/SF\\
$^{290}114$  & $^{286}112$  &0.075    &0.000&-0.034  &0.008  &8.84 &-4.35729&3.55858  &4.04192 &3.97597 &4.21954& SF\\
\cline{1-13}

$^{308}122$  & $^{304}120$  & 0.000  & 0.000 & 0.000  & 0.000 &15.29 &14.0826&-9.11754 &-9.43036 &-8.60751 &-7.87666 &$\alpha$\\
$^{304}120$  & $^{300}118$  & 0.000  & 0.000 & 0.000  & 0.000 &13.55&6.56723&-6.61842 &-6.7198 &-6.05901 &-5.45516  &$\alpha$\\
$^{300}118$  & $^{296}116$  &-0.011  & 0.000 & 0.000  & 0.000 &12.51 &0.475263 &-5.06451&-5.08261 &-4.51391 &-4.00775 &$\alpha$\\
$^{296}116$  & $^{292}114$  & 0.000    & 0.000& 0.000  & 0.000  &11.18 &  -4.28234 &-2.62826&-2.48818 &-2.08208 &-1.68853& SF\\
\cline{1-13}

$^{310}122$  & $^{306}120$  & 0.000  & 0.000 & 0.000  & 0.000 &13.12 &10.0421&-5.30718 &-5.27712 &-4.71008 &-4.10911 &$\alpha$\\
$^{306}120$  & $^{302}118$  & 0.000 &0.000 &0.000  &0.000 &14.27 &2.49992& -7.98553 &-8.15542 &-7.36643 &-6.76532  &$\alpha$\\
$^{302}118$  & $^{298}116$  &0.000  &0.000 &0.000  &0.000 &12.62 &-3.61905 &-5.37571& -5.35945&-4.74243 &-4.26469 &$\alpha$\\
$^{298}116$  & $^{294}114$  &0.000    &0.000&0.000  &0.000  &11.13 & -8.40377& -2.55752 &-2.38829 &-1.96011 &-1.60276& SF\\
\cline{1-13}

$^{312}122$  & $^{294}120$  &-0.407  &0.000 &-0.003  &0.024 &12.74 &5.28887 &-7.11981& -4.46731&-3.92695 &-3.37854 &$\alpha$\\
$^{308}120$  & $^{290}118$  & 0.000 &0.000 &0.000  &0.000 &12.97 &-2.28013 &-5.60586    & -5.56696&-4.92757 &-4.41772  &$\alpha$\\
$^{304}118$  & $^{282}116$  &0.000  &0.000 &0.000  &0.000 &13.39&-8.42604 &-6.95934& -7.02386&-6.26235 &-5.7824 & SF\\
\cline{1-13}

$^{314}122$  & $^{310}120$  &-0.407  &0.000 &-0.003  &0.024 &12.58  &-0.176908 & -5.55731 &-4.13277 &-3.58664 &-3.07914 &$\alpha$\\
$^{310}120$  & $^{306}118$  &0.000  &0.000 &0.000  &0.000 &11.28 &-7.77269 &-1.82474& -1.5367&-1.14684 &-0.759457  & SF\\

\cline{1-13}

$^{316}122$  & $^{312}120$  & -0.416 &0.000 &0.010  &0.019 &12.19 &-6.35502 & -5.77157   & -3.2428&-2.72925 & -2.2746& SF\\

\cline{1-13}
\end{tabular}
\end{table}

In table 2, all the chains stop when spontaneous fission is reached. Some steps in the process show that either alpha decay or spontaneous fission can happen, this is due to the half-life values being quite close to each other for both the decay modes.

The plots for individual isotopes illustrates the competition between fission and alpha decay during the chain process.

\begin{figure}[H]
\center
\includegraphics[height=6.8cm, width=8.5cm]{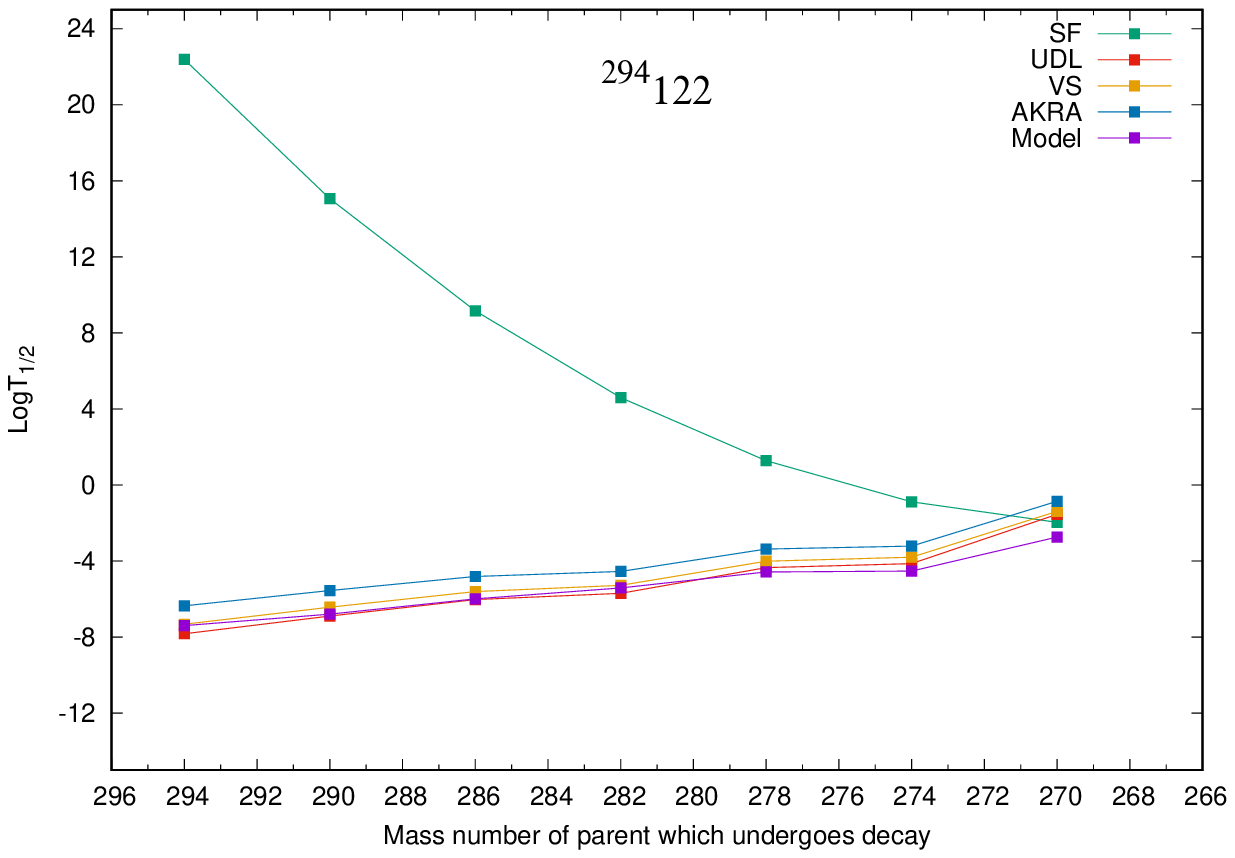}
\includegraphics[height=6.8cm, width=8.5cm]{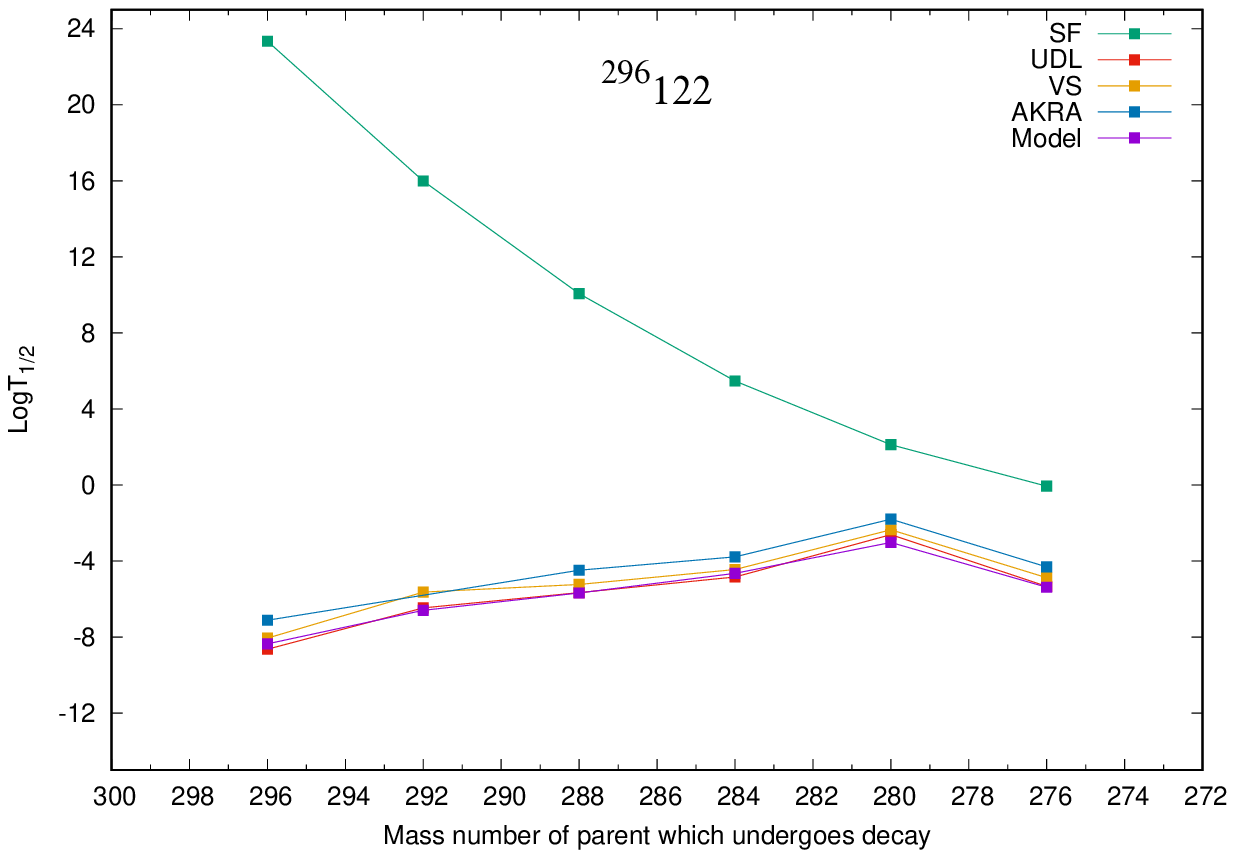}
\end{figure}

\begin{figure}[H]
\center
\includegraphics[height=6.8cm, width=8.5cm]{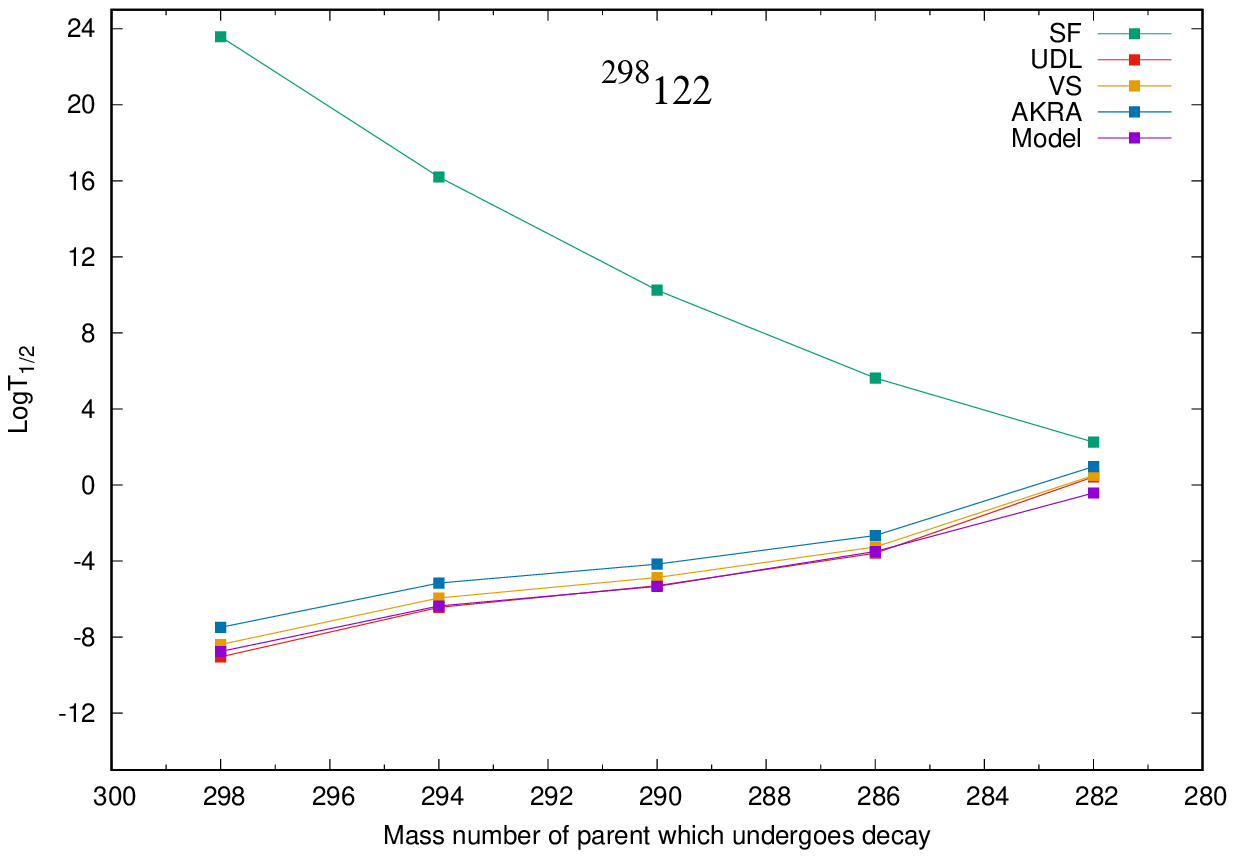}
\includegraphics[height=6.8cm, width=8.5cm]{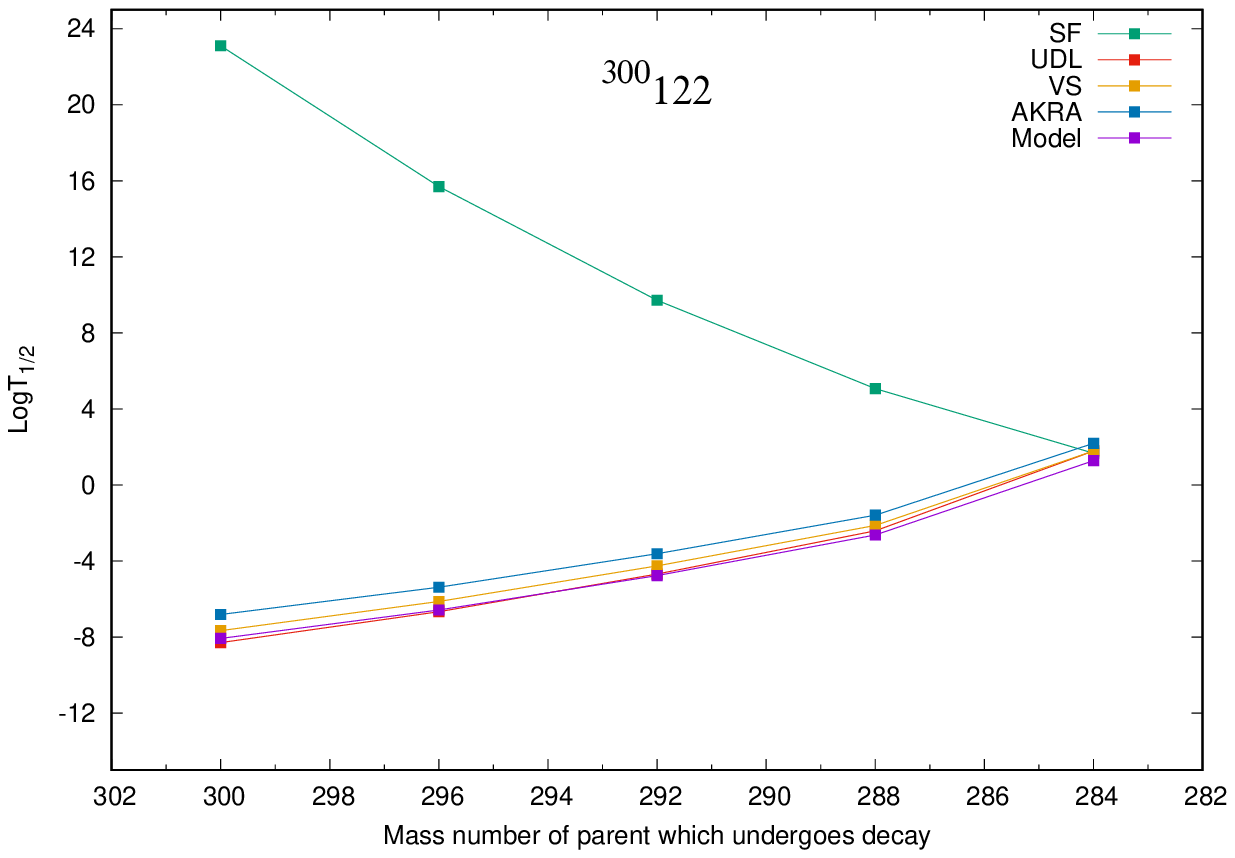}
\end{figure}

\begin{figure}[H]
\center
\includegraphics[height=6.8cm, width=8.5cm]{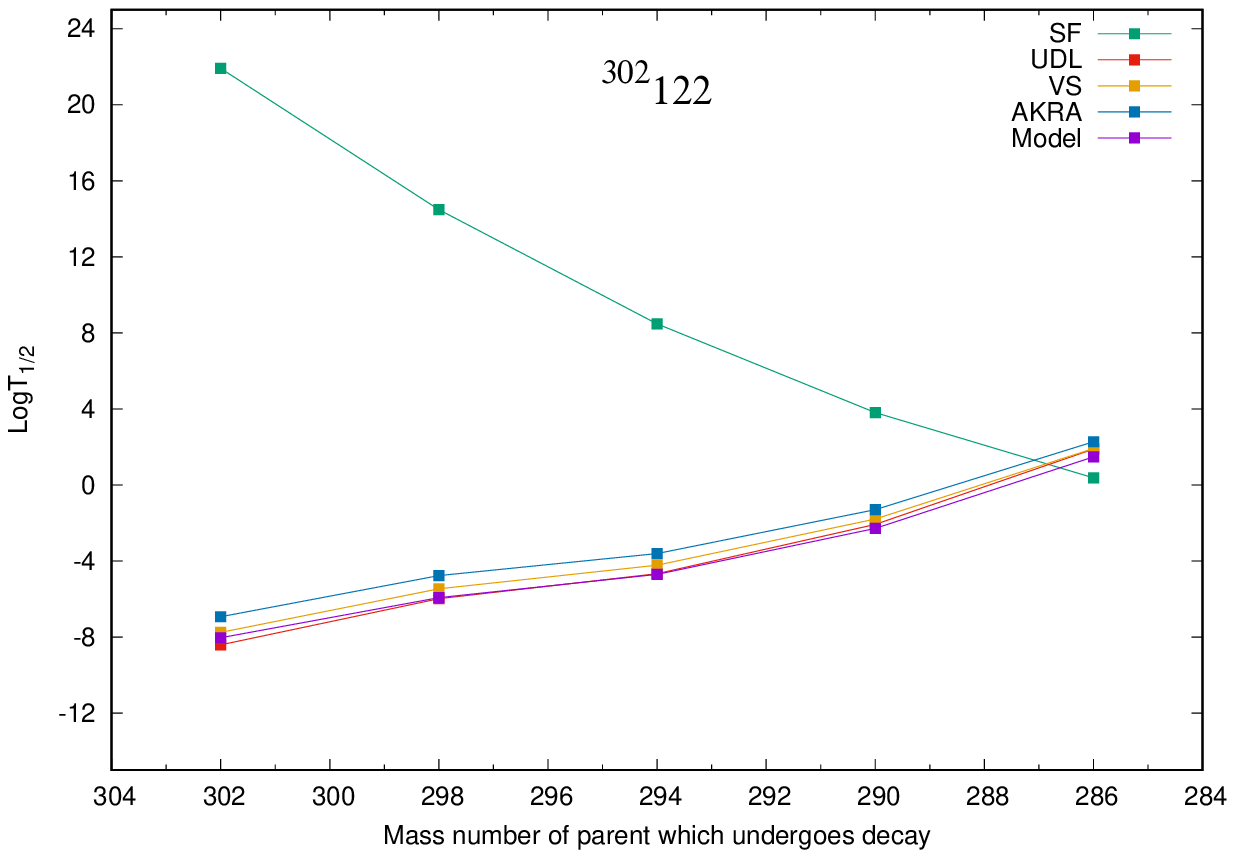}
\includegraphics[height=6.8cm, width=8.5cm]{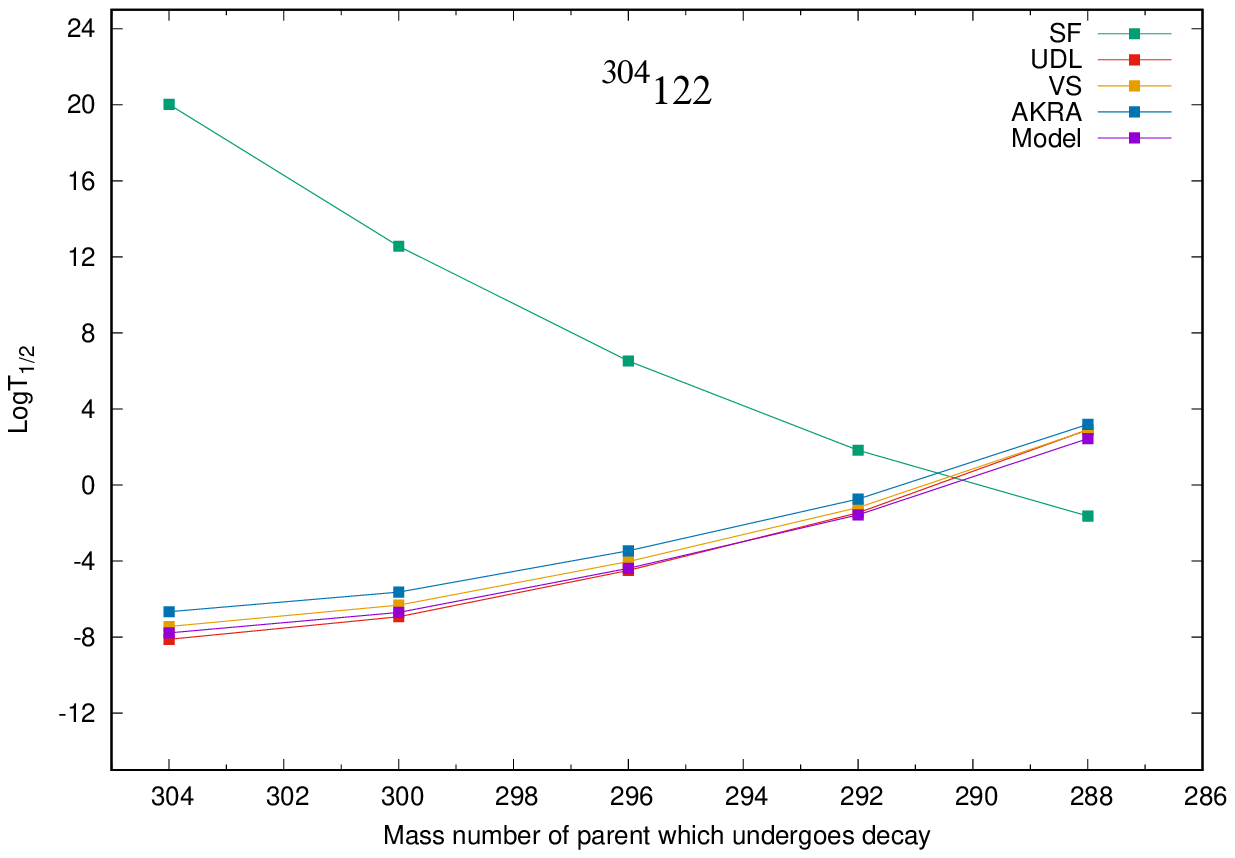}
\caption{Alpha decay and Spontaneous fission half-lives of various isotopes of $Z=122$}

\end{figure}

\begin{figure}[H]
\center
\includegraphics[height=6.8cm, width=8.5cm]{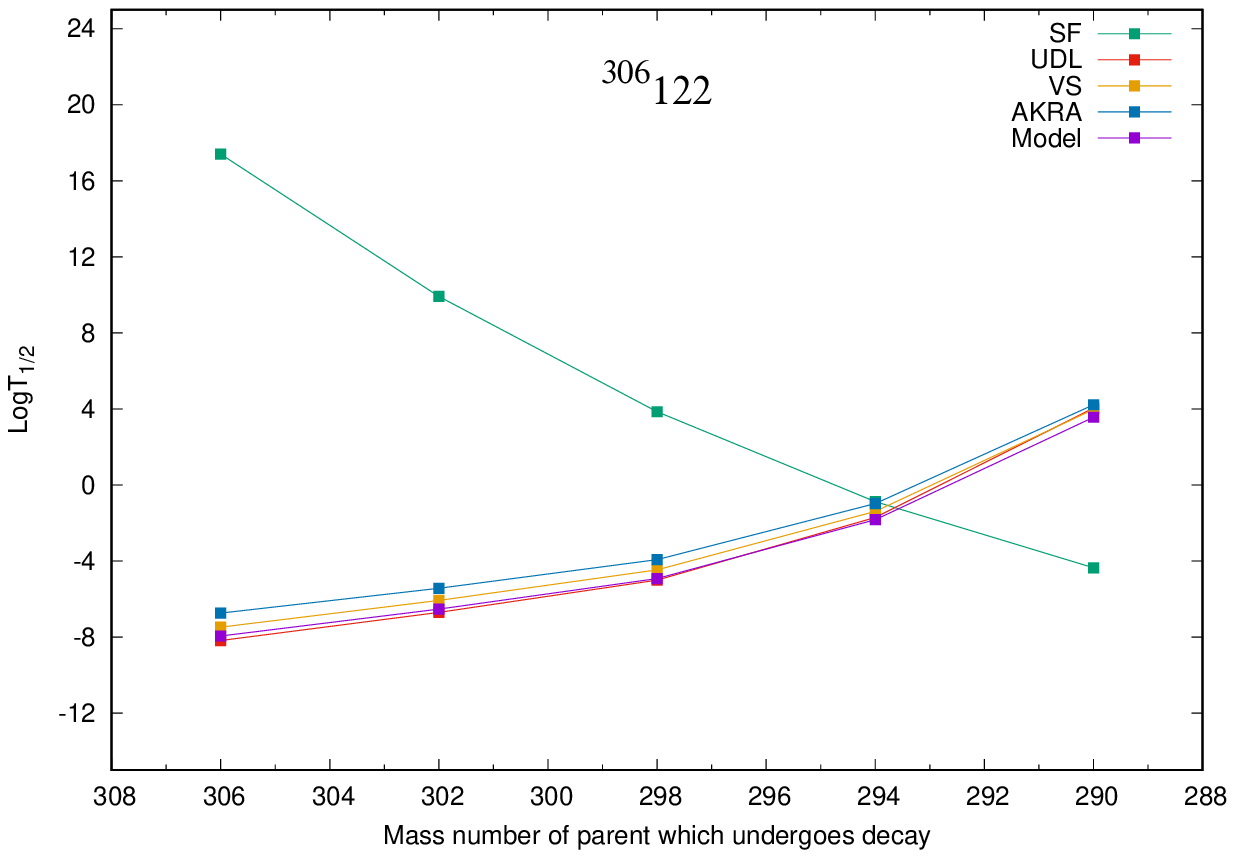}
\includegraphics[height=6.8cm, width=8.5cm]{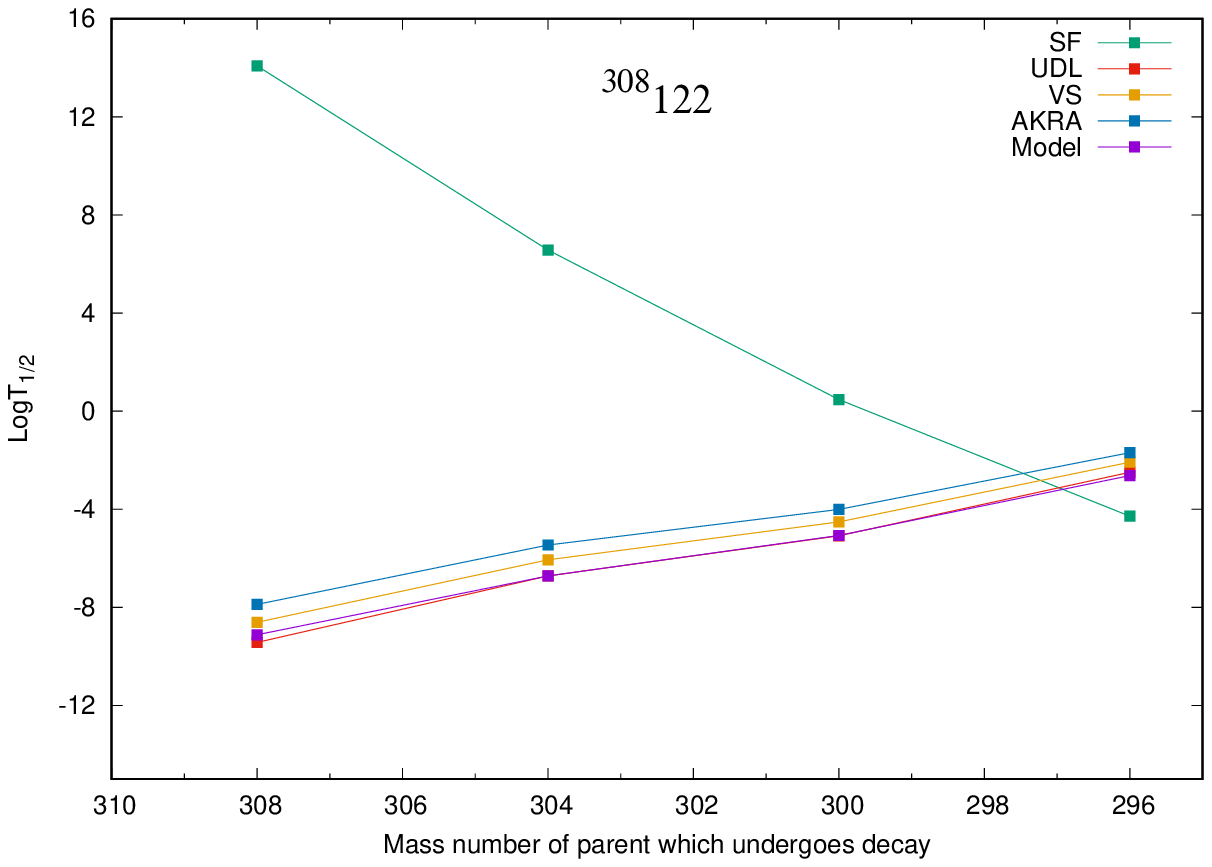}
\end{figure}
\begin{figure}[H]
\center

\includegraphics[height=6.8cm, width=8.5cm]{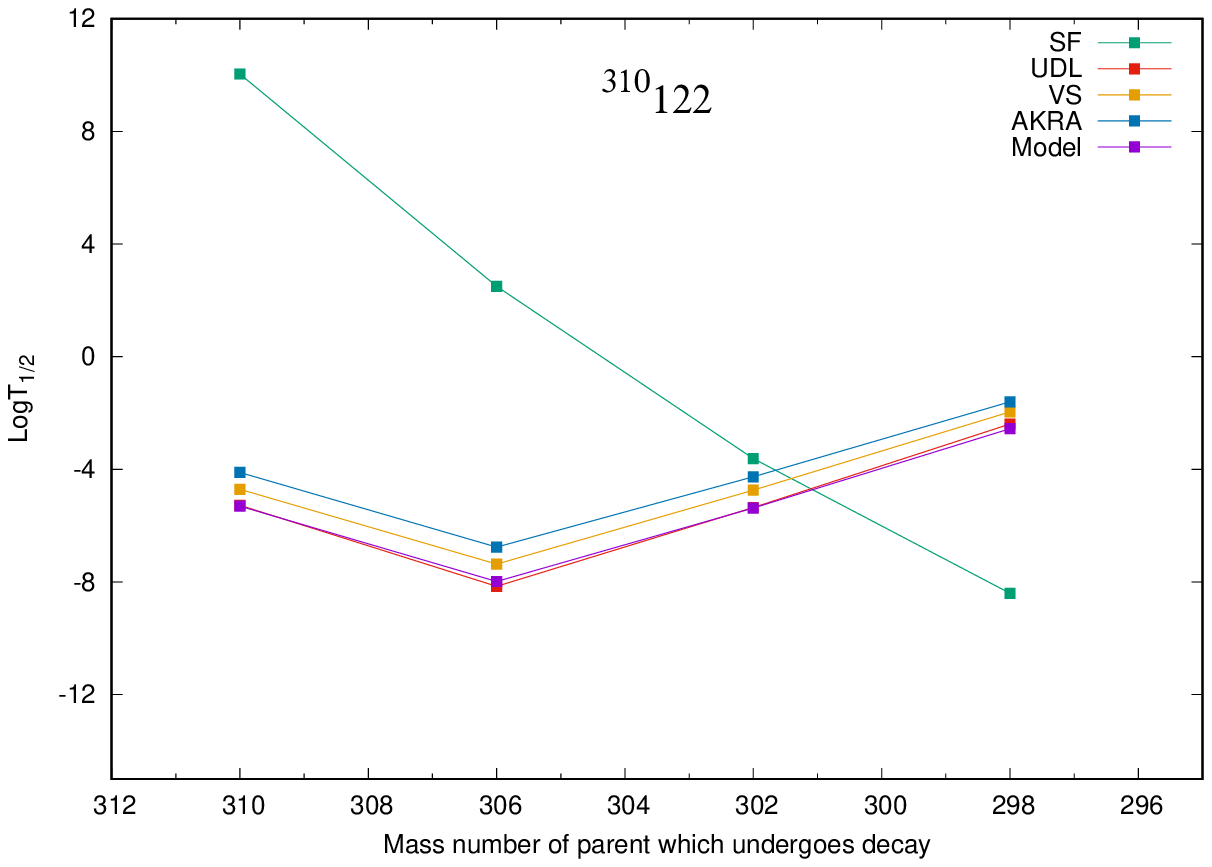}
\includegraphics[height=6.8cm, width=8.5cm]{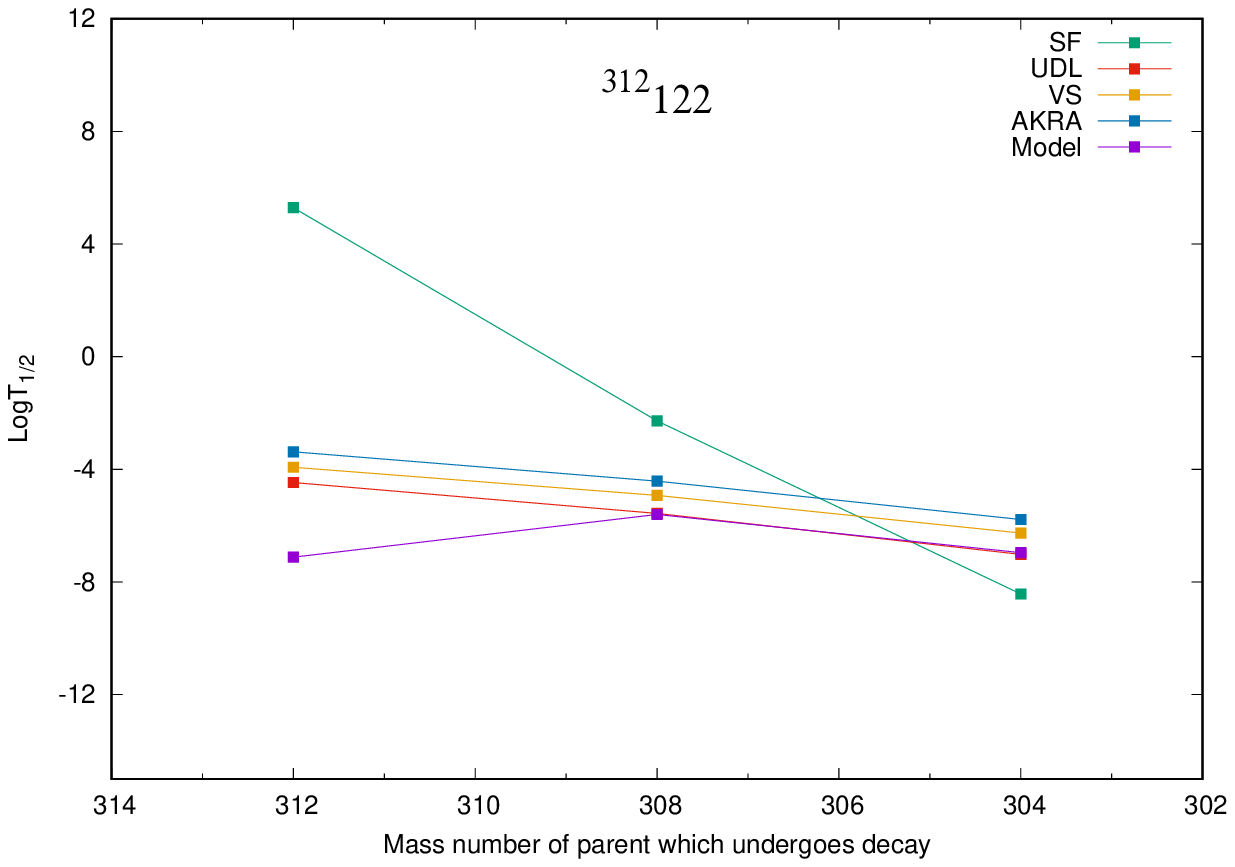}
\includegraphics[height=6.8cm, width=8.5cm]{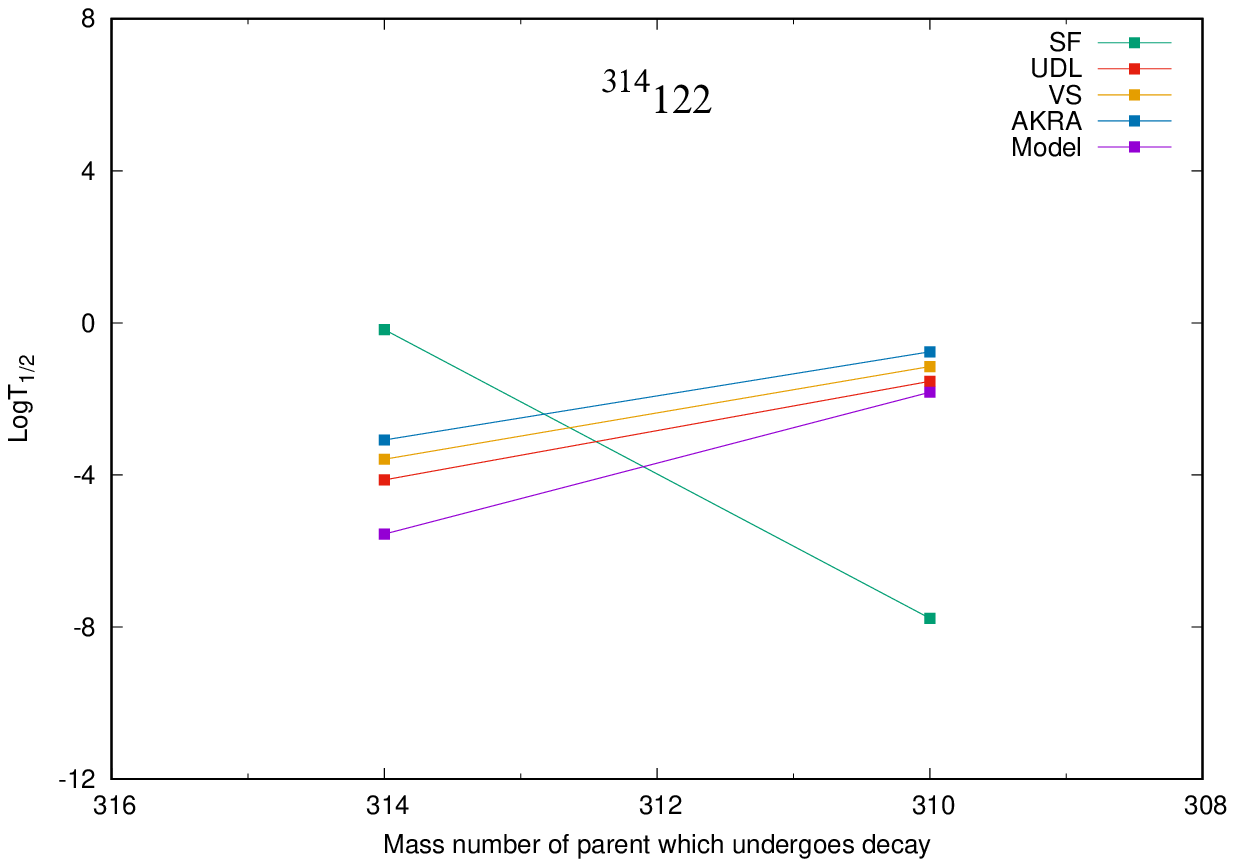}

\caption{Alpha decay and Spontaneous fission half-lives of various isotopes of $Z=122$}
\end{figure}

In the plot the spontaneous fission half-life decreases during the chain process and when becomes less than alpha decay half-life the decay chain stops, this is seen as a crossing of fission curve with alpha decay curve. Some points where spontaneous fission half-life close to alpha decay half-life, both modes of decay are possible. The crossing of fission curve is not seen in some isotopes, this is due to next step in the chain being unfavorable to Q value of process going negative.

\begin{figure}[H]
\center

\includegraphics[height=8cm, width=14cm]{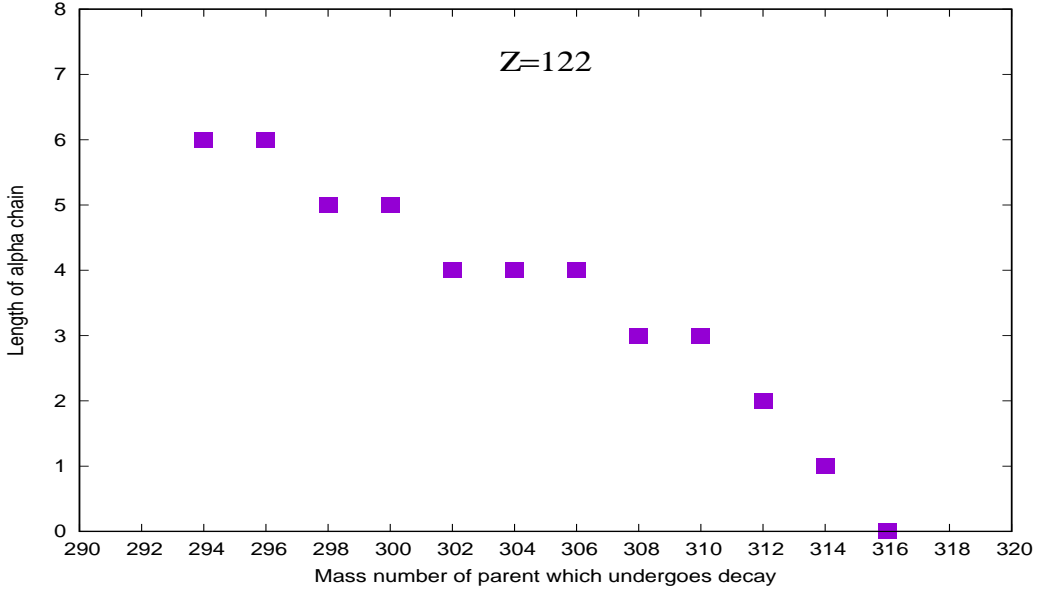}

\caption{Alpha decay chain length of even isotopes of $Z=122$ before onset of spontaneous fission}
\end{figure}

\section{Summary $\&$ Conclusion}

Our model is based on the famous Woods-Saxon potential, which very well describes the nuclear potential. In addition to it for Coulomb potential, the deformation and orientation effects of nuclei are considered which have been emphasized in recent years for theoretical evaluation of decay half lives. The form of the centrifugal term is also given importance, when the centrifugal term is taken in the total potential and WKB integral is performed over the 1D radial coordinate it requires the use of Langer modified term, which involves using $(l+ \frac{1}{2})^{2}$ instead of $l(l+1)$. The calculated half-life seems to be well in agreement with the values found from the phenomenological formula of alpha decay. 

The alpha decay half-life value is compared with the spontaneous fission half-life value, to find the possible mode of decay. Most isotopes of $Z=122$ is expected to emit alpha chains and decay rapidly. The spontaneous fission half-life value decreases with each successive step of the alpha decay chain, whereas the alpha decay half-life is seen to increase with each step of the alpha decay chain. Our work suggests that $\alpha$ chain consisting of a sequence of 6 alpha decays will be seen from $^{294}122$ and $^{296}122$. Five alpha decays from $^{298}122$, and 4 alpha decays from $^{300}122$, $^{302}122$,$^{304}122$, and $^{306}122$ will happen. Smaller chains are also expected to occur. We find that alpha chain with 3 alpha decays will occur in $^{308}122$ and $^{310}122$, whereas 2 alpha decay from $^{312}122$. There will be only a single alpha decay from $^{314}122$. There will be no alpha chain seen in $^{316}122$.\\

The no. of decays in the decay chains along with the $Q$ values will be the unique signature associated with each of the isotopes. We hope that the results of our work will be of help to the future experimentalist in choosing which of the $z=122$ isotope/ isotopes they would like to go for synthesis in a lab. Our results will be helpful also in the identification of the isotope synthesized in the experiment by tracking the unique decay signature reported here.\\
 
\section*{Acknowledgement}
The authors would like to thank Department of Nuclear Physics, University of Madras for providing all the research and infrastructure facilities.

\end{document}